\documentclass[10pt, paper=a4, UKenglish]{article}
\usepackage{graphicx}
\def\Title#1{\begin{center} {\Large #1 } \end{center}}
\def\Author#1{\begin{center}{ \sc #1} \end{center}}
\def\Address#1{\begin{center}{ \it #1} \end{center}}

\newcommand\pubblock{\rightline{\begin{tabular}{l} Proceedings of the CTD/WIT 2019\\ \pubnumber\\
         \pubdate  \end{tabular}}}

\newenvironment{Abstract}
	{\begin{quotation}
 	\begin{center} 
		\large ABSTRACT
	\end{center}\bigskip
		\begin{normalsize}
			}
			{
		\par 
		\end{normalsize}
	\end{quotation}
	}

\newenvironment{Presented}{\begin{quotation} \begin{center} 
             PRESENTED AT\end{center}\bigskip 
      \begin{center}\begin{large}}{\end{large}\end{center} \end{quotation}}

\def\beq{\begin{equation}}
\def\eeq#1{\label{#1}\end{equation}}
\def\eeqn{\end{equation}}

\def\beqa{\begin{eqnarray}}
\def\eeqa#1{\label{#1}\end{eqnarray}}
\def\eeqan{\end{eqnarray}}

\let\bar=\overbar

\def\Dslash{\not{\hbox{\kern-4pt $D$}}}
\def\dslash{\not{\hbox{\kern-2pt $\del$}}}

\def\msb{{\bar{\ssstyle M \kern -1pt S}}}

\usepackage{color}
\usepackage{lineno}
\usepackage{subfig}
\usepackage{hyperref}
\usepackage{amsmath} 
\usepackage{diagbox}

\newcommand\pubnumber{PROC-CTD19-007}

\newcommand\pubdate{\today}

\def\affiliation{
Physikalisches Institut \\
University of Heidelberg, Germany}

\newcommand{\conference}{Connecting the Dots and Workshop on Intelligent Trackers (CTD/WIT 2019)\\
Instituto de F\'isica Corpuscular (IFIC), Valencia, Spain\\ 
April 2-5, 2019}

\usepackage{fancyhdr}
\definecolor{mygrey}{RGB}{105,105,105}
\begin{document}

\large
\begin{titlepage}
\pubblock

\vfill
\Title{Triplet Track Trigger for Future High Rate Experiments}
\vfill

\Author{Tamasi Kar, Danilo E. F. de Lima, Andr\'e Sch\"oning \& Jike Wang}
\Address{\affiliation}
\vfill

\begin{Abstract}
The hadron-hadron based Future Circular Collider (FCC-hh) is a project with the goal to collide proton-proton beams at a center of mass energy of $\sqrt{s} = $100 TeV with a bunch crossing rate of 25 ns. 
Some of the major challenges that the FCC-experiments have to tackle are the very large number of pile-up events $\sim$1000 and the data processing, namely the reduction of the huge data rate of 1 - 2 PBytes/s whilst keeping the signal efficiencies of interesting processes high.
Therefore, smart triggering concepts are needed that not only allow for a significant reduction of pile-up and rate but also provide high signal acceptance and purity.

In this proceeding, one such concept of a triplet track trigger based on Monolithic Active Pixel Sensors (MAPS) is presented for a generic detector geometry.
It is demonstrated that the triplet pixel layer design allows for a very simple and fast track reconstruction already at the first trigger level, providing excellent track reconstruction efficiencies and very high purity at the same time.
Based on a full Geant4 simulation tracking performance studies are presented for a full-scale triplet pixel detector, i.e. three closely spaced pixel layers at sufficiently large radius, in an FCC-like detector environment. 
Results obtained for different triplet layer design parameters are compared.
\end{Abstract}

\vfill

\begin{Presented}
\conference
\end{Presented}
\vfill
\end{titlepage}
\def\thefootnote{\fnsymbol{footnote}}
\setcounter{footnote}{0}
%

\normalsize 


\section{Introduction}
\label{intro}

Several accelerator projects are under study for the post High Luminosity era \cite{futureColliders} to increase the discovery potential for new physics at both the high energy and intensity frontier. 
One of the many projects that are being studied is the 
FCC-hh which aims to collide proton beams at $\sqrt{s} =$ 100 TeV at a bunch crossing rate of 25 ns \cite{FCC}. 
Such high energy hadron collisions will not only allow high precision measurement of Higgs boson properties, but also allow search for rare processes with high sensitivity. 

One of the main challenges that such high rate experiments would have to deal with is a very large pile-up\footnote{pile-up here refers to the number of proton-proton collisions per bunch crossing of 25 ns} ($\sim$1000) in a luminous region of $\sim$10 cm along the beam direction (z-axis).
Additionally, in the high rate experiments like ATLAS and CMS, the very first stage of the trigger system is based on the energy deposited in the calorimeters.
In order to clearly distinguish the hard proton-proton collisions from the soft ones, vertex information with a resolution of better than a tenth of an mm is required. 
Availability of tracks at the very first stage of a trigger system will, therefore, prove to be very crucial in suppressing pile-up by a significant amount. 
Section~\ref{TTT_concept} introduces one such concept of a track trigger based on detector triplet consisting of  MAPS, called as the Triplet Track Trigger (TTT). It relies on a very simple and fast track reconstruction algorithm, detailed in Section~\ref{TTRA}.
The $hh \rightarrow bbbb$ physics channel has been used to study the performance of the TTT for an FCC-hh like detector environment, and the results obtained for different triplet layer design parameters are compared in Section~\ref{results} for two different average number of pile-up interactions configurations. 
\section{Triplet Track Trigger concept}
\label{TTT_concept}
Track reconstruction in high rate experiments is very challenging and computationally expensive due to a large number of possible combinations from the hits in different layers.
Consequently, this technically limits the access to the reconstructed track parameters at a very early stage of the trigger system in high rate experiments like ATLAS and CMS.

The concept of Triplet Track Trigger was proposed for reconstructing tracks in real time at such high rate experiments \cite{Aschoening1}, and uses a very simple and fast track reconstruction algorithm.
A detector triplet is defined as a stack of three closely spaced, highly granular pixel detector layers (preferably MAPS due to their low production cost and the possibility to instrument large areas) placed at large radii ($> $40 cm). 
When a charged particle passes through such a detector triplet in a uniform magnetic field \textbf{B}, its trajectory is circular in the x-y plane and a straight line in the s-z plane, see Figure~\ref{fig:TTTR}.
A detector triplet allows for an easy reconstruction of TTT tracks and provides enough redundancy to efficiently reduce combinatorial background.
In addition, an excellent spatial resolution of pixels allows determination of z-vertex with very high precision. 
The TTT concept relies on the assumption that all particles originate from the beamline\footnote{beamline constraint assumes that the particles originate from (0, 0, $z_0$)} or have a small transverse impact parameter, O(1 cm), if they originate from secondary decays.
The large radius of a detector triplet not only ensures a good curvature resolution and hence momentum resolution of the reconstructed tracks, but also reduces the impact parameter dependence of secondary particles in momentum reconstruction.
The track reconstruction algorithm used is detailed in the following section.
\section{Triplet Track Reconstruction Algorithm}
\label{TTRA}
\begin{figure}[!htb]
  \centering
  \subfloat[]{\includegraphics[width=0.30\linewidth]{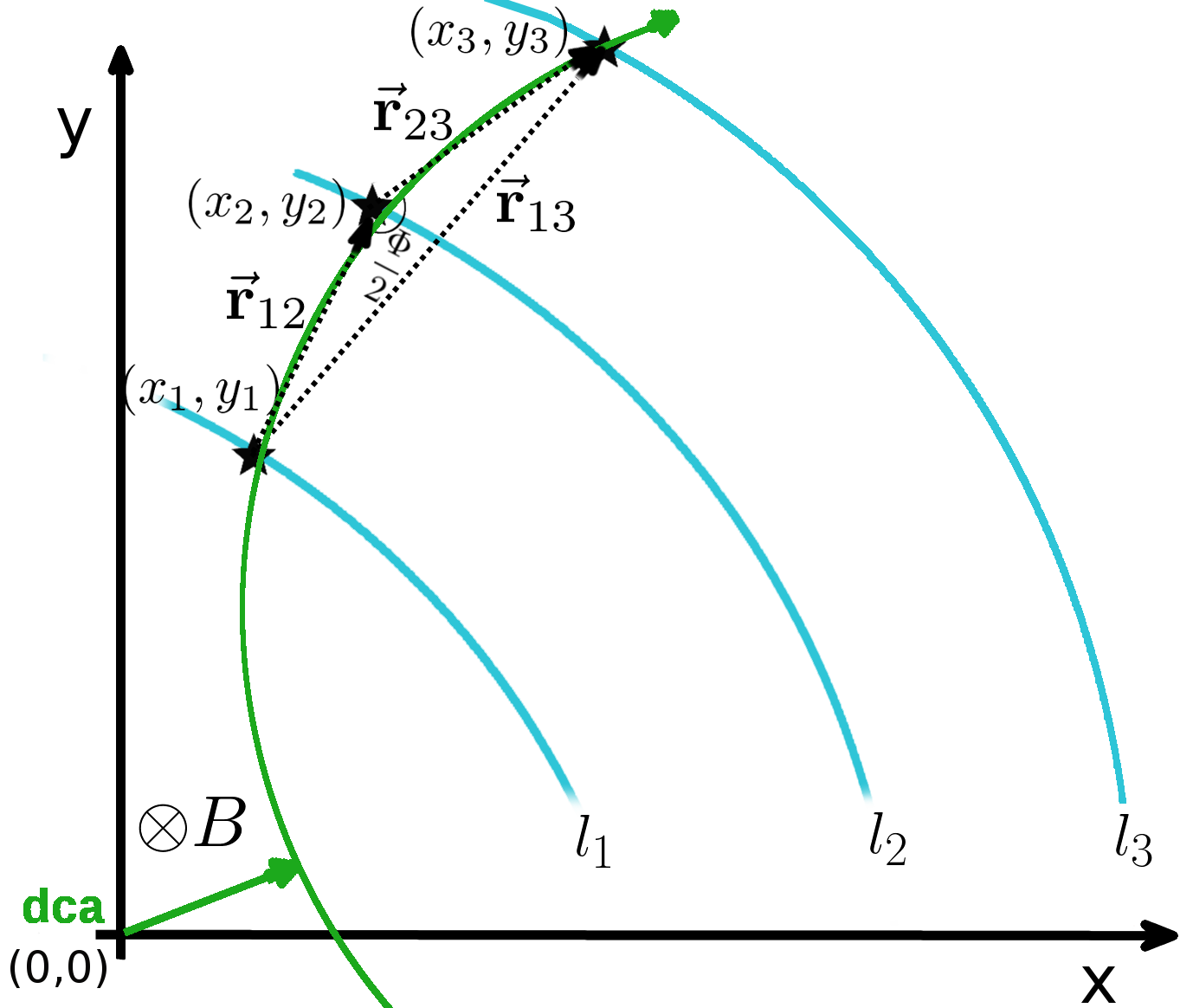}}
  \qquad
  \qquad
  \qquad
  \subfloat[]{\includegraphics[width=0.30\linewidth]{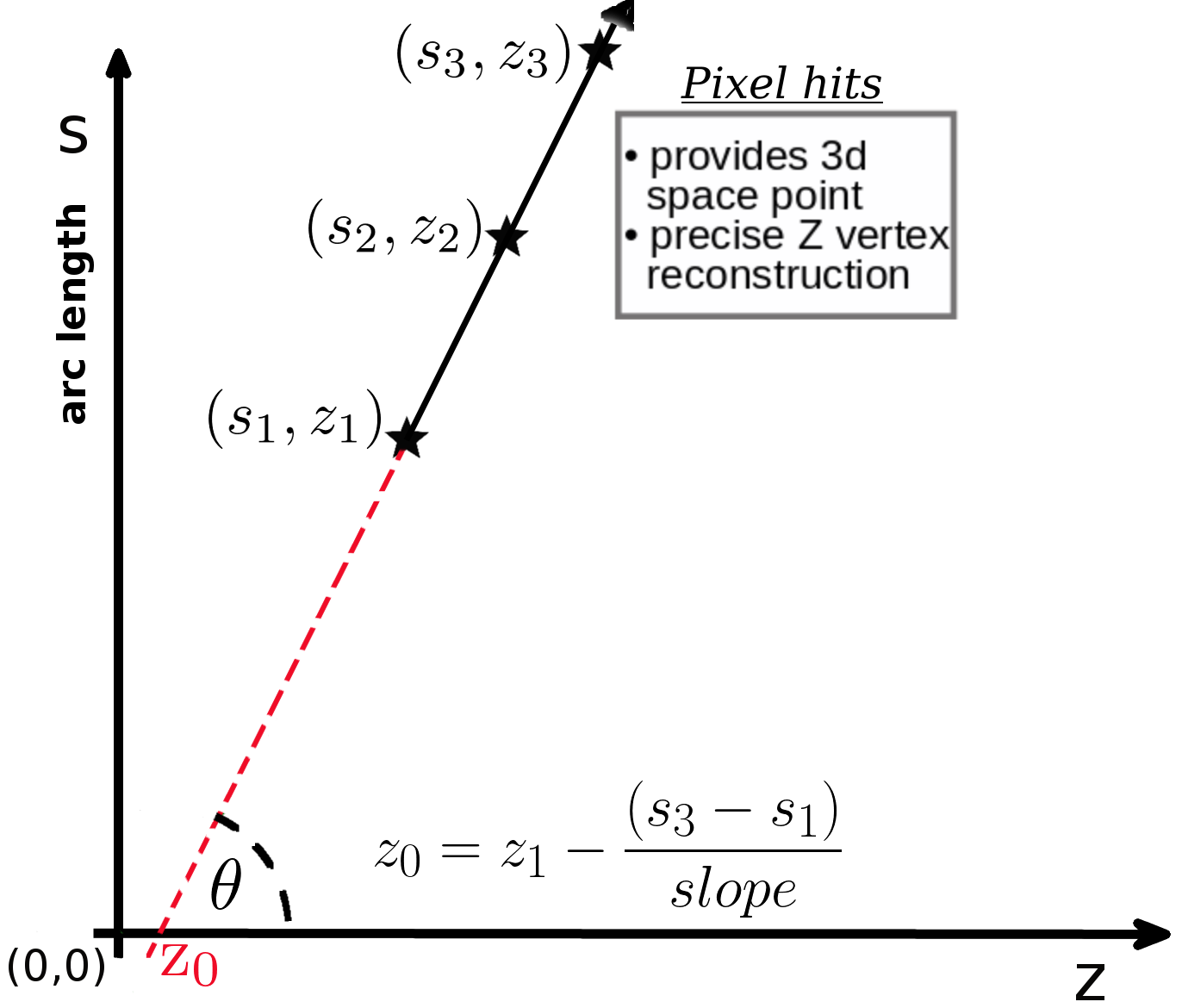}}
  \caption{(a) reconstruction of a charged particle's track in the transverse plane, in a uniform magnetic field \textbf{B}. (b) z-vertex reconstruction of this track in the longitudinal plane.}
  \label{fig:TTTR}
\end{figure}

Consider a TTT in a uniform magnetic field \textbf{B} comprising layers $l_1$, $l_2$ and $l_3$ placed at a radius R, a section of which is shown in Figure~\ref{fig:TTTR}a.
A charged particle passing through it is detected with coordinates $(x_i, y_i, z_i)$ in the $i^{th}$ layer, where $i = 1, 2, 3$. Let $\phi_i = \tan^{-1}(y_i/x_i)$ be the measured azimuthal angle corresponding to the $i^{th}$  hit, its trajectory is reconstructed using the following three steps: triplet hit selection,  triplet track reconstruction, final selection cuts.
\subsection{Triplet hit selection}
\label{THS}
In this step, a search window in $\Delta{z}-\Delta{\phi}$ is defined w.r.t the hit co-ordinates in $l_3$ to search for hit candidates in $l_1$. Tracks can already be reconstructed at this stage using the beamline constraint, see Section~\ref{TTR}. 
Since the trajectory between $l_1$ and $l_3$ is almost a straight line for small gap sizes and large radii of the TTT, a cut on the residua of the hit in the middle layer is applied both in the x-y and the s-z planes:
\begin{align}
|d\phi_{2}| &= |\phi_2 - 0.5*(\phi_1 + \phi_3)| < d\phi_{2,cut} \\
|dz_{2}| &= |z_2 - 0.5*(z_1 + z_3)| \: \: < dz_{2,cut} * sin(\theta)^{n_z}
\end{align}
where a polar angle dependence of $\theta$ is included assuming multiple coulomb scattering in $l_2$. The hit combinatorial problem is largely reduced at this stage and this makes a triplet design special compared to the tracker designs based on doublets.
The cut parameters $d\phi_{2,cut}$, $dz_{2,cut}$ and $n_z$ are optimized using simulated minimum bias particles for a given geometry.
\subsection{Triplet track reconstruction}
\label{TTR}
The triplet hit candidates selected above are used to determine the track parameters analytically with two independent methods by solving for a circle and a straight line in the x-y \& the s-z planes, respectively.
At first, 
the track curvature $\kappa$, and hence the transverse momentum $p_t$, is determined using the hits in $l_1$, $l_2$ and $l_3$.
If $\Phi$ is the angle subtended by the chord joining the first and the third hit at the center of the circular trajectory traversed  by a particle, then
\begin{align}
sin(\Phi/2) &= \frac{\vec{\mathbf{r}}_{23} \times \vec{\mathbf{r}}_{12}}{|\vec{\mathbf{r}}_{23}|*|\vec{\mathbf{r}}_{12}|} \\
\kappa = \frac{1}{\textrm{radius}} &= \frac{2*sin(\Phi/2)}{|\vec{\mathbf{r}}_{13}|} \\
p_t &= \frac{0.3*B}{\kappa}
\end{align}
where $\vec{\mathbf{r}}_{ij}$'s are the position vectors pointing from hit i to j, see Figure~\ref{fig:TTTR}a.\\ 
In the s-z plane, the z-vertex, $z_0$ and $\theta$ are extrapolated from the straight line joining the hits in $l_1$ and  $l_3$, see Figure~\ref{fig:TTTR}b.
The remaining track parameters dca\footnote{dca is in the transverse plane, the shortest distance between the circular trajectory and the beamline (0, 0, $z_0$), see Figure~\ref{fig:TTTR}a} and $\varphi$\footnote{$\varphi$ is the initial direction of a particle in the x-y plane. It is orthogonal to the line joining the center of the circle and the point of closest approach} are determined by knowing the center of the above circle.

The same set of equations are used to determine the above track parameters with the beamline constraint.
In this method, the track parameters in the transverse plane can be re-calculated using the hits in $l_1$ and $l_3$ alone, as a pseudo-hit $(x_0, y_0) = (0, 0)$ is used as the third hit. 
Notice that, dca is zero by construction and the lever arm $|\vec{\mathbf{r}}_{03}|$, determining the curvature is much larger compared to $|\vec{\mathbf{r}}_{13}|$ in the previous method.
As a result, $\kappa$ and hence the momentum resolution determined is much more precise with the beamline constraint.
\subsection{Final selection cuts}
As a final step, a few more selection requirements are imposed on the precise parameters obtained with beamline constraint, based on the luminous region ($|z_0| < 10cm$) of the colliding proton beams, the acceptance ($|\eta| < 1.6$) of the TTT and the $p_t$ threshold (2 GeV/c).
A significant number of wrongly reconstructed tracks are further rejected by making a consistency check on the curvature values determined using the two independent methods described in Section~\ref{TTR}. 
If $d\kappa$ is the difference between the curvatures determined with the above two methods, then a momentum consistency cut can be defined as:
\begin{align}
|d\kappa| < n * \sigma_{\kappa}
\end{align} 
where n is an acceptance cut in units of the standard deviation and $\sigma{\kappa}^2 = K_{Hit}^2 + \kappa^2 * K_{MS}^2$ is the curvature uncertainty which is determined by a constant hit uncertainty term and a $\kappa$ dependent multiple scattering term \cite{Aschoening2}.

\section{Results}
\label{results}
\begin{figure}[!htb]
  \centering
  \subfloat[]{\includegraphics[width=0.35\linewidth]{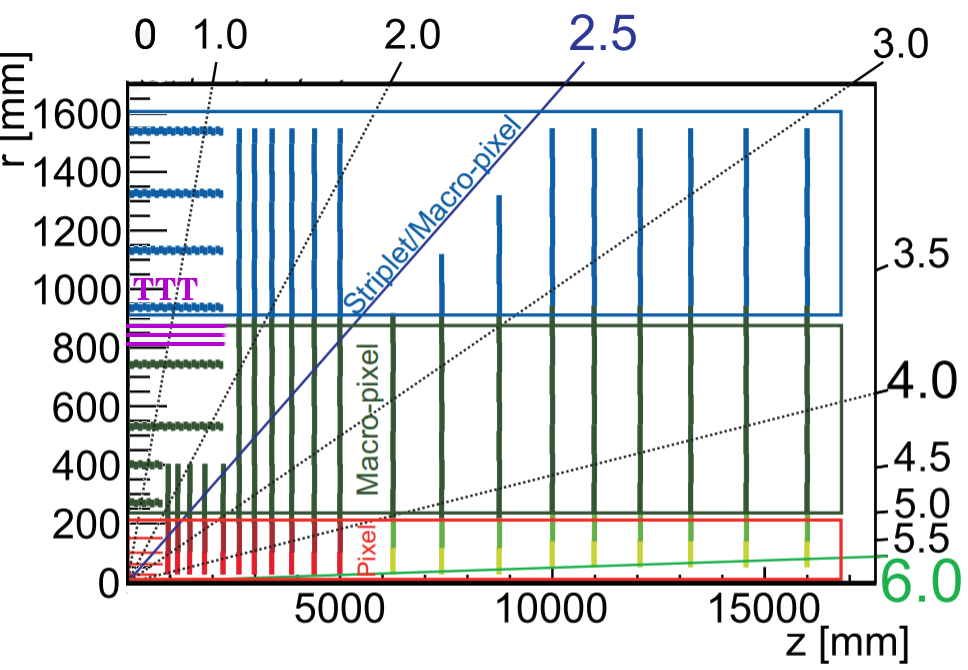}}
  \qquad
  \qquad
  \subfloat[]{\includegraphics[width=0.52\linewidth]{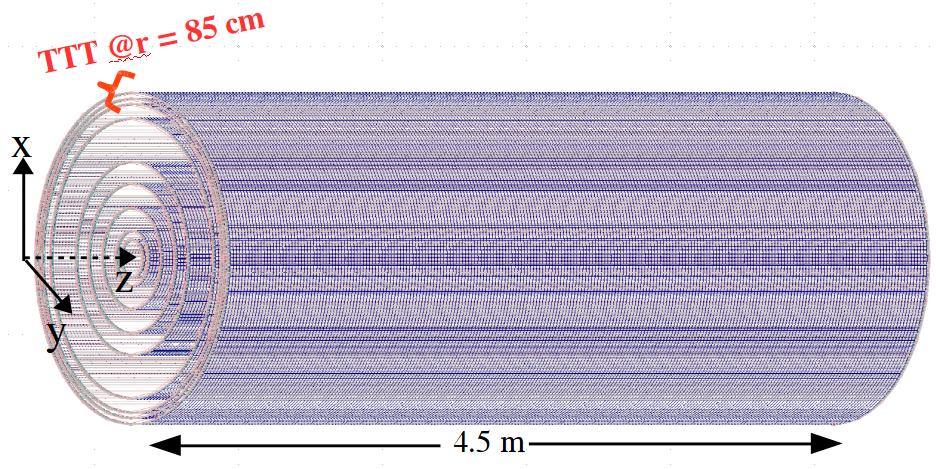}}
  \caption{(a) TTT at a radius of 85 cm for FCC-hh tracker layout and (b) detector geometry in Geant4 for the central tracker }
  \label{fig:TKlayout}
\end{figure}

To study the tracking performance of the TTT in a high pile-up environment, a detector geometry similar to the baseline tracker layout \cite{FCC} of the FCC-hh (Figure~\ref{fig:TKlayout}a) is implemented in Geant4 (Figure~\ref{fig:TKlayout}b) with the three layers located at a radius of about 85 cm.
For simplicity, all the layers in this design have a length of 4.5 m and a relative radiation length of 2\%/layer.
The granularity of the pixels is chosen to be $40 \times 40 \mu m^{2}$ and detailed simulations are performed for three different gap sizes of the TTT, namely 20, 40, 50 mm and for a pile-up of 200 and 1000.
Tracks were reconstructed using the previously mentioned reconstruction algorithm in a uniform magnetic field of 4 T. 
\begin{figure}[!htb]
  \centering
  \subfloat[]{\includegraphics[width=0.325\linewidth]{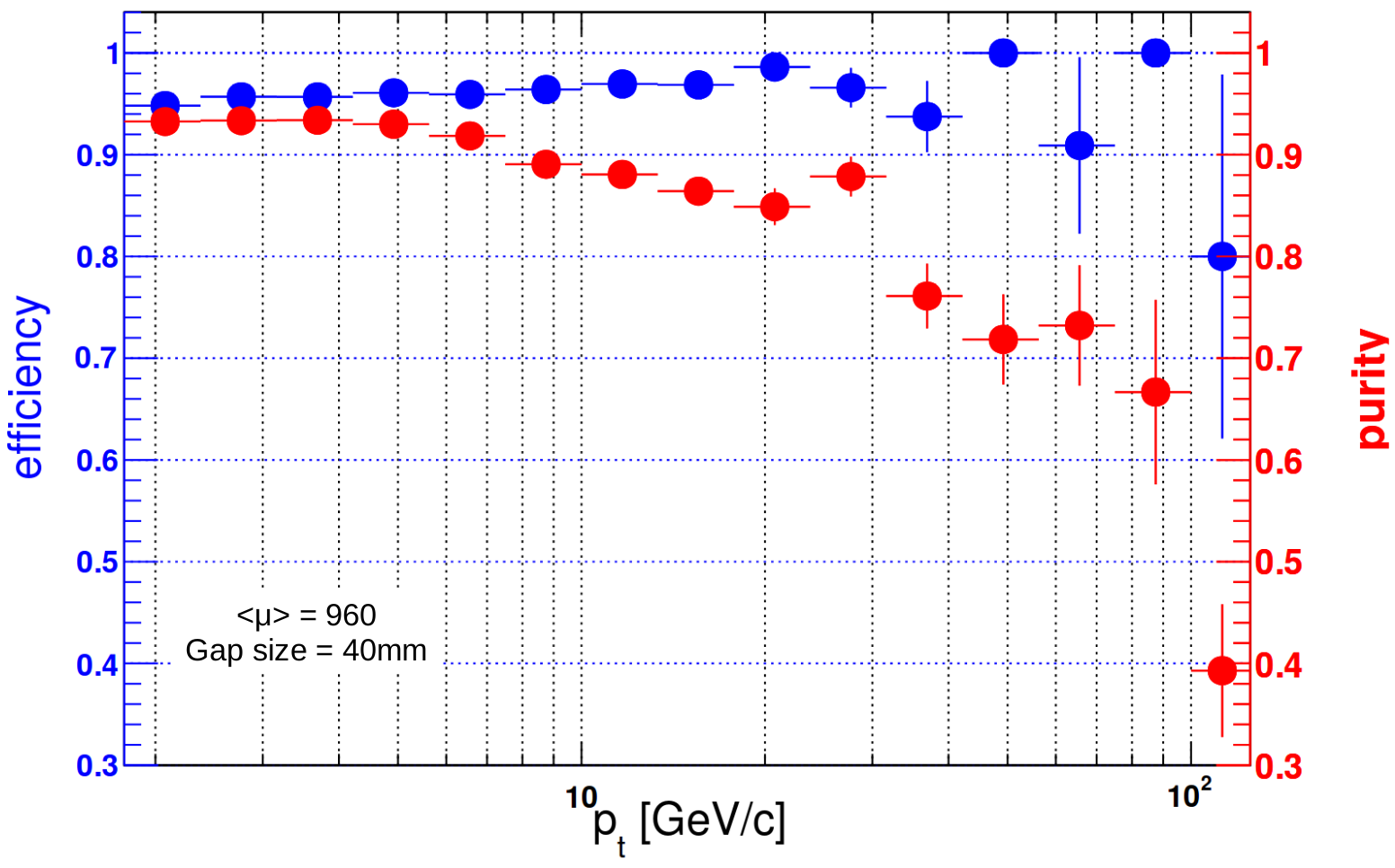}}
  \quad
  \subfloat[]{\includegraphics[width=0.30\linewidth]{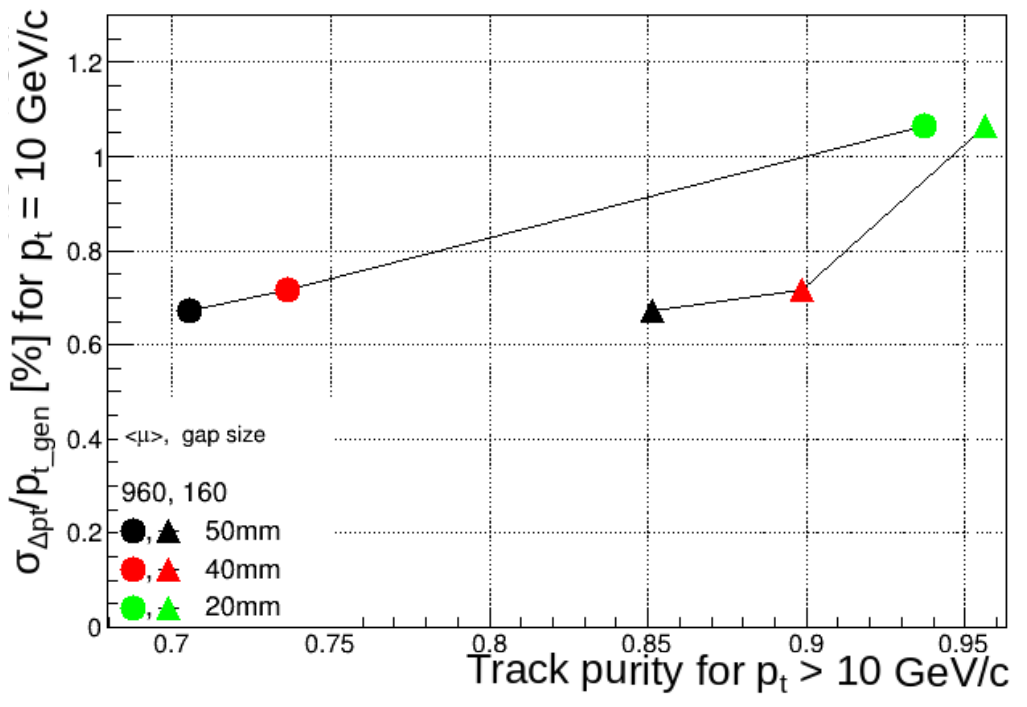}}
  \quad
  \subfloat[]{\includegraphics[width=0.30\linewidth]{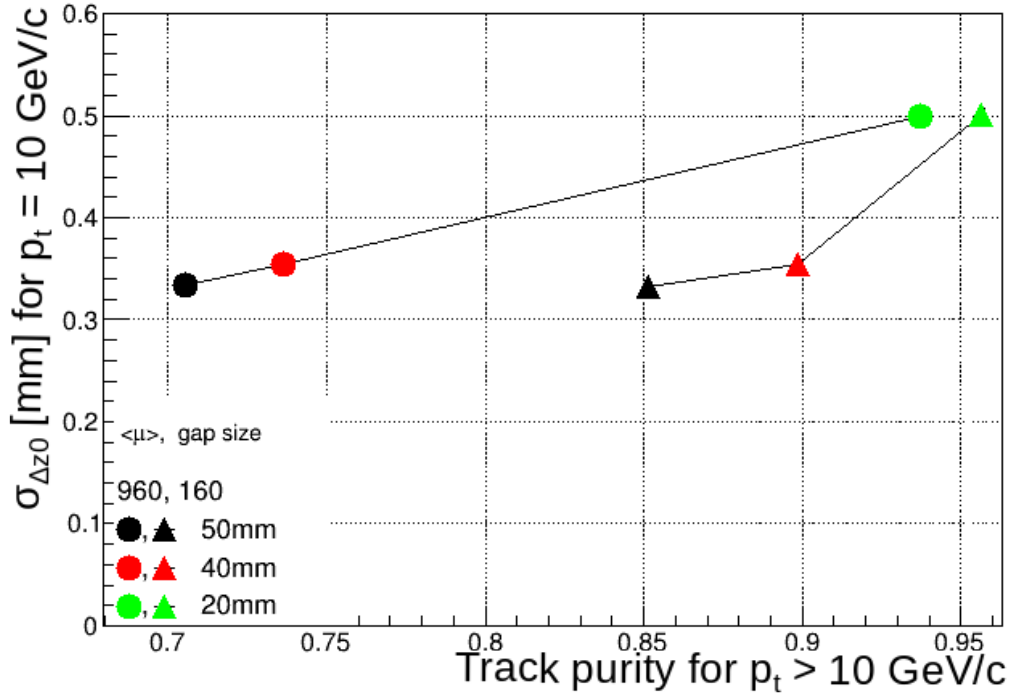}}
  \caption{(a) shows the track reconstruction efficiency and purity in $hh \rightarrow bbbb$ as a function of transverse momentum in pile-up 1000 events for a triplet layer gap of 40 mm.
Relative $p_t$ resolution (b) and $z_0$ resolution (c) for $p_t$ = 10 GeV/c pions as a function of track purity in $hh \rightarrow bbbb$ for $p_t > 10$ GeV/c is shown for three different gap sizes of the TTT in pile-up 200 and 1000 respectively.}
  \label{fig:results}
\end{figure}
The results shown below were produced using the $hh \rightarrow bbbb$ sample with the colliding proton beams having a center of mass energy of 13 TeV\footnote{Simulated samples of $hh \rightarrow bbbb$ with $\sqrt{s} = $100 TeV and pile-up 1000 were not ready at the time of the conference}. 
Tracks are reconstructed with an efficiency $>95\%$ and with a high track purity even at a very high pile-up of 1000, see Figure~\ref{fig:results}a.
In Figures~\ref{fig:results}b--c, the relative transverse momentum resolution and $z_0$ resolution of 10 GeV/c pions are plotted as a function of track purity in $hh \rightarrow bbbb$ integrated over $p_t >$ 10 GeV/c.
A momentum resolution of better than $1\%$ at 10 GeV/c is achievable for a gap size of TTT $>20$ mm. 
$z_0$ is reconstructed with sub-mm precision, thereby allowing a significant pile-up reduction at a very early stage in the trigger system.
For a given pile-up configuration, the $p_t$ and the $z_0$ resolutions increase as a function of the gap size of the TTT, while the track purity decreases.
Track purity degrades more rapidly with the gap size of the TTT with increasing pile-up. Therefore, a gap size between 20 to 40 mm would be optimal for the TTT.
\section{Summary and Outlook}

The concept of triplet track trigger employs a very simple and fast track reconstruction algorithm and can be implemented in hardware, such as an FPGA.
It will therefore, provide fast access to tracks in a trigger system allowing smart selection and significant pile-up reduction.
It was shown that TTT has a very good tracking performance even at a pile-up of 1000 and is able to resolve z-vertices with sub-mm precision.

The TTT simulation environment will be used to perform trigger performance studies. 
The focus will be set on processes (e.g. $hh \rightarrow bbbb$) which are difficult to trigger by the calorimeters alone. 
It is planned to study single- and multi-high pt track topologies as well as track-jets as signatures for multi-jet events and to evaluate achievable trigger rate reductions for different signatures.
Full simulation and performance studies of TTT discs in the forward eta region are foreseen as part of future studies. 




\end{document}